\documentstyle[12pt,aasms4]{article}

\def\ga{\gtrsim}
\def\la{\lesssim}
\def\tfm{TFM }
\def\rtfm{RTFM }
\def\eg{{\it e.g.,\ }}
\begin{document}
\title{Bifurcation, efficiency, and the role of injection in shock 
acceleration with the Bohm diffusion}
\author{M. A. Malkov}
\affil{ Max-Planck-Institut f\"ur 
Kernphysik, D-69029, Heidelberg, Germany; \\
malkov@boris.mpi-hd.mpg.de \\
} 

\begin{abstract}
The efficiency and uniqueness of the diffusive shock acceleration is 
studied on the basis of the novel kinetic solutions.  These solutions 
obtained earlier (paper I) selfconsistently describe a strong coupling of 
cosmic rays with the gas flow.  They show that the dependence of the 
acceleration efficiency upon physical parameters is critical in nature.  In 
this paper we investigate a steady acceleration in the parameter space 
formed by the injection rate \( \nu \), the upper cut-off momentum \( p_1 
\) and the Mach number \( M \) while the flow compression $R $ serves as an 
order parameter.  We determine a manifold of all possible solutions in this 
parameter space.

To elucidate the differences between the present kinetic results and 
the well known two-fluid predictions we particularly focus on the \(\nu \to 
0, \, M \to \infty \) limit where the two-fluid model suffers from 
especially serious closure problems and displays an `unphysical' behavior.  
We show that in contrast to the two-fluid model three different solutions 
occurs also for arbitrarily large \( M \) provided that \( p_1 \) is 
sufficiently high.  The three solutions appear together only if the 
injection rate $\nu $ lies between two critical values, \( \nu_1 < \nu < 
\nu_2\).  For \( \nu < \nu_1(M,p_1) \) only the inefficient solution is 
possible.  For \( \nu > \nu_2(M,p_1) \) only the efficient solution with a 
very high cosmic ray production rate occurs.  On the basis of the obtained 
bifurcation surface \( R(\nu, p_1) \) we consider the limit \( p_1 \to 
\infty , \, \nu \to 0 \) which completely uncovers the long debating 
anomalies of the two-fluid model.

The constructed steady state manifold that, at least partially is an 
attractor of a time dependent system, allows us to speculate on the 
nonstationary acceleration.
\end{abstract}


\section{Introduction}
The question, how efficient the diffusive shock acceleration may be, arose 
naturally when the first test particle calculations of this process became 
available (Krimsky 1977; Axford, Leer \& Skadron 1977; Bell 1978; Blandford 
\& Ostriker 1978; see \cite{dru}; \cite{bla:eich} and \cite{jel91} for a 
review).  This is because the backreaction of accelerated particles (cosmic 
rays (CRs) in the astrophysical context) on the shock structure is very 
strong and leads normally to a significant increase of the compression 
ratio.  Such an accelerating shock should therefore be thought of as a 
strongly nonlinear dynamical system with a pronounced selforganization.  
Neither the particle spectrum nor the hydrodynamic flow structure can be 
calculated independently.  Furthermore, since the diffusion length of 
particles increases with momentum, particles with higher momenta sample 
longer parts of the shock transition.  This makes any kind of moment 
description very difficult.  
However, the first essentially nonlinear calculations of this
acceleration process were performed within the hydrodynamic approach.
 \subsection{Success and limitations of the two-fluid model}
The above arguments suggest that the problem is kinetic in nature,
which makes the usage of any fluid theory for describing the
acceleration process questionable.  At the same time quite a deep
insight can be gained from simple moment equations.  The two-fluid
model (TFM) introduced by Axford Leer \& Skadron (1977) and Drury \&
V\"olk (1981) (DV, hereafter) treats the thermal and CR populations as
separate fluids coupled only through the hydrodynamic equations.  The
main effect of this coupling is a deceleration of the inflowing gas in
front of the shock by the pressure gradient of counterstreaming CRs
accelerated at the shock and, as a result, an enhancement of the total
shock compression,  the multiplicity of solutions and a much higher
acceleration efficiency.  Unfortunately, the underlying particle
distribution implies the pressure divergence  and is underdetermined in
some other ways (see \eg \cite{dru}; \cite{abp}; \cite{kj90}, and
below).
\subsection{Renormalization of the two-fluid model} 
A renormalization procedure to overcome the above ultraviolet
divergence has been suggested recently by Malkov \& V\"olk (1996)
(MV96, hereafter).  This theory produces basically the same two-fluid
hydrodynamics except the renormalized CR specific heat ratio $\Gamma$
instead of the usual \( \gamma_{\rm c}\) which results from the
losses.  Under the assumption $\gamma_{\rm c} =5/3$ upstream (which
automatically implies that far upstream $\Gamma=5/3$ as well, for the
case of reacceleration considered in MV96), $\gamma_{\rm c} =4/3$
downstream, and in the limit $p_1 \to \infty$, the renormalized
two-fluid model (RTFM) produces a solution qualitatively similar to
that of the unrenormalized theory.  In the case of $\gamma_{\rm c}
=4/3$ upstream, when also \(\Gamma \) decreases $\Gamma < \gamma_{\rm
c}$, the results change dramatically.  Namely, the shock compression
$r$ becomes much larger than the usual unrenormalized result $r=7$ (for
strong shocks).  What happens is a very fast increase of the losses
with decrease of $\gamma_{\rm c}$ which rises  the compression that
even tends to infinity (for infinitely large Mach numbers when, in
addition, one takes the maximum possible spectral slope $q =3r/(r-1)$
at the upper cut-off).  This regime was not (and could not be) explored
in MV96 since such a high compression shock requires a detailed
information about $\gamma_{\rm c},\, \Gamma $, and $\bar\kappa$, the
spectrum averaged CR-diffusivity across the shock transition.  This
would practically be equivalent to the full kinetic solution.  To
obtain such a solution was one of the main objectives of a companion
paper (Malkov 1997a, paper I).  Further motivations of the present paper
will be outlined in the next subsection.

To conclude this subsection we note that the assumption \( \gamma_{\rm
c} \approx 4/3 \) upstream is precisely what the kinetic solution
obtained in paper I suggests for the case of injection in contrast to
the case of reacceleration, \( \gamma_{\rm c} = 5/3 \) considered in
MV96.  Moreover, $\gamma_{\rm c}  $ is smaller upstream than
downstream, again opposite to the reacceleration case. Finally, the
\rtfm results in the injection case are unacceptably sensitive to the
values of $\gamma_{\rm c}  $ and \( \Gamma \)  that are not known to
the required extent when the kinetic solution is not available.  This
makes the moment approach especially restrictive for describing namely
the injection triggered acceleration process.  As in paper I we
concentrate here on this, certainly more interesting and at the same
time more difficult case.
\subsection{`Pathological' limits of the \tfm}
There exists another
difficulty of the \tfm that has already been criticized in the
literature (\eg \cite{jel91}). Namely, the shock modification in a
steady state occurs while particles are constantly injected at some
rate, are then accelerated, and disappear eventually through the upper
cut-off or downstream.  Once the injection is somehow eliminated, the
CR-dominated (or efficient) steady-state solution cannot be justified
physically, and the ordinary gas shock remains the only solution
possible. The TFM, however, still produces a CR-dominated shock that
even becomes completely smooth beyond a certain Mach number, \( M > M_1 \).  
In fact, it represents a fast-mode shock in a two-fluid hydrodynamics 
associated with the tenuous high pressure CR-fluid (\cite{ptus}).  
Since the number density of the CRs \( n_{\rm c} \) is irrelevant in the 
TFM, solutions that have a finite CR pressure (\( P_{\rm c} > 0\)) are 
formally permitted without any injection (\( n_{\rm c} =0\)).  Moreover, 
for an arbitrarily small nonzero injection rate there exists a critical 
Mach number \( M_2 < \infty \) above which this efficient solution is the 
only one the \tfm can offer.  

Clearly, it is difficult to judge the acceleration efficiency because the 
injection rate is, as a rule, very small and the two-fluid system, on the 
other hand, does not behave adequately when injection vanishes.  The 
question, however, is whether the consideration of this limit within the 
\tfm is admissible.  The answer is definitive not.  Indeed, a fully kinetic 
steady state solution obtained in paper I revealed the following nonlinear 
response of the system to a weak injection of thermal particles.  First of 
all, this response depends not so much on the injection itself but rather 
on the parameter \( \Lambda_1=\nu/\delta \equiv \nu p_1/p_0 \), where \( 
\nu \) and \( p_0 \) are the injection rate and injection momentum, 
respectively, and \( \delta \ll 1 \).  Furthermore, the effect of shock 
modification completely disappears as \( \Lambda_1 \to 0 \) (it becomes 
practically insignificant in an abrupt manner already at \( \nu 
/\sqrt{\delta} \la 1 \)).  Only in the other extreme, \( \Lambda_1 \to 
\infty \) (more precisely, when \( \Lambda_1 \gg M^{3/4} \), see also 
subsection \ref{crit:inj}) a {\em unique} solution that is indeed injection 
insensitive appears which is quite in the spirit of the TFM.  
Quantitatively, this solution can be very different from the respective 
\tfm solution but for a different reason which is related to the 
subshock smoothing.  The TFM is of course incapable of describing the 
dependence of its solution on \( \Lambda_1 \) since it implies \( 
p_1=\infty \), i.e.  \( \Lambda_1 =\infty \) already on the derivation 
level; even if \( \nu \) is set to zero afterwards the physically correct 
behavior of the solution with vanishing injection cannot be recovered since 
\( \Lambda_1 \) remains infinite.  Therefore it is useless to expect from 
the \tfm a correct behavior at \( \nu \to 0 \) and to criticize this model 
for the lack of it.  This is beyond its validity range.

The most dramatic consequence of this (\( p_1 \to \infty \))
degeneracy is the subshock smoothing (\( r_{\rm s} \equiv 1 \)).  The
solution becomes enormously different from the kinetic solution that
does not pass through the point \( r_{\rm s} =1 \) just because of this
fact.  Why this is so, has been explained in paper I and we shall look
at this problem from a different perspective in subsection
\ref{crit:inj}.

Perhaps the most direct explanation why the kinetic solution differs so 
strongly from its hydrodynamic counterpart is the singular character of the 
underlying perturbation problem in the small parameter \(\delta \equiv 
p_0/p_1 \ll 1 \).  No matter how small it is, the efficient kinetic solution 
with \( \delta = 0 \) (and, hence, the \tfm solution) is fundamentally 
different from that with \( \delta > 0 \) (see paper I).

The parameters that the \tfm usually operates on are the Mach number \( M 
\) and the injection rate \( \nu \) (or the seed particle pressure in the 
case of reacceleration).  As we argued, this is not enough to describe 
consistently what is going on in the steady nonlinear shock acceleration.  
The \rtfm introduces an additional parameter needed, \(p_1 \), and accounts 
of the losses at \( p = p_1 \) but then it lets \( p_1 \to \infty \), \( 
\nu \) fixed.  That means \( \Lambda_1 \equiv \nu /\delta \to \infty \), 
and therefore the results are again insensitive to \( \nu \) when \( \nu 
\to 0 \) since the critical information about \( \Lambda_1 \) is lost, 
exactly as in the TFM.  We emphasize that it is the parameter \( \Lambda_1 
\) that regulates primarily the budget of energetic particles at a shock, 
not \( \nu \) alone.  That is why, for example, the subshock completely 
vanishes in the \tfm as well as in the \rtfm beyond a certain \( M =M_1\) 
even for \( \nu \to 0 \); the more important parameter \( \Lambda_1 \) 
remains infinite which effectively corresponds to the situation with a very 
strong injection.  

If we allow for time dependence on the kinetic level of description and
assume a slower than in the Bohm limit momentum dependence of the CR
diffusion coefficient, completely smooth \tfm stationary solutions will
exist even with zero injection and with no seed particles upstream.
This phenomenon has been explained by Drury (1983)-- the high energy
particles may be considered as those injected in the past and being
then continuously accelerated at the shock.  Unfortunately the smooth
\tfm solutions can tell us very little about a steady acceleration of
CRs out of the thermal upstream plasma -- if there are no preexisting
CRs, there are no such solutions.

Since these \tfm solutions are essentially time dependent on the
kinetic level (see \eg \cite{fg87}, \cite{kj90} and \cite{dvb} for
relevant discussions), we may infer that once a natural cut-off \( p_1
< \infty\) exists and is reached, this acceleration regime will be
disrupted.  Indeed, these solutions correspond to the acceleration of
CRs injected in the past whose number density virtually decreases
(although being irrelevant in the \tfm which in fact admits these
pseudo-steady solutions) while the CR pressure remains approximately
constant, being determined simply by the ram pressure of the inflowing
gas.  When particles start to leak through the upper cut-off the CR
pressure decreases as well and the system relaxes to the ordinary gas
shock which is the only steady state solution without injection. In
general, the above \tfm quasi-stationary acceleration scenario implies
a rather low production of CRs (in terms of their number, not the
energy density) because it operates only on initially injected
particles and suppress further injection as soon as this solution is
set up.  Much more productive would be solutions that allow for
permanent losses. This would mean, in fact, the propagation of
high-energy CRs into the shock surroundings which decouples them from
the gas flow with their replenishment due to the permanent injection at
the subshock.  But these solutions are essentially kinetic and, as we
emphasized, fairly different from the \tfm solutions.

As it was demonstrated in paper I, given the injection rate three
different solutions are possible for sufficiently high \( M \) and \(
p_1 \), Figure 1.  However, only the most efficient solution with the
highest compression ratio has been considered in detail.  Accordingly,
only a first critical injection, i.e.  the injection rate \( \nu =
\nu_1 \) above which this solution exists along with the two other
solutions has been calculated.  The calculation of a second critical
injection that requires an inspection of the two remaining solutions
and above which the efficient solution is the only possible, is one of
the subjects of the present paper.  We consider the entire manifold of
stationary solutions and in this context the three above-mentioned
solutions are simply its subsets.

In the next section we briefly review the physical formulation of the 
problem and discuss our strategy of a unified description of all the three 
solutions.  In Sec.3 we obtain both the efficient and inefficient solutions 
from the integral equation derived in paper I and consider their matching 
in an intermediate range.  In Sec.4 we describe the solution space as a 
whole and calculate the critical injections. We conclude this section 
with implications of its results for the \tfm.  Further, in Sec.5 we 
speculate upon possible scenarios of time dependent acceleration on the 
basis of the emerged bifurcation picture.  Sec.  6 discusses the results 
and some of their most evident consequences for  calculations of the 
acceleration efficiency in real astrophysical objects.
\section{Kinetic solution}\label{ks}
The standard formulation of the problem of diffusive shock acceleration 
includes the diffusion-convection equation for the high energy particles 
constrained by conservation of the fluxes of mass and momentum (see \eg 
\cite{dru}; \cite{bla:eich}).  The physical situation that we shall 
consider throughout this paper is described in paper I and we quote only 
the key elements below.
\subsection{Overview of the physical formulation} 
We assume that a strong CR-modified shock propagates in the positive \( x 
\)-direction and in the shock frame of reference the steady state velocity 
profile of the gas is specified as follows: \( U(x)=-u(x) \) for \( x \ge 
0 \), \( 
u(0+)=u_0 \); \( u(\infty)=u_1 \).   The equation that 
describes the isotropic part of particle distribution $g(x,p) $ at 
sufficiently 
high momenta is the diffusion-convection equation (Parker 1965; Gleeson \& 
Axford 1967) which in the region \( x \ge 0 \) takes the form
\begin{equation}
	\frac{\partial}{\partial x} \left( u g + \kappa(p) \frac{\partial 
g}{\partial x}\right) =\frac{1}{3} \frac{du}{dx} p\frac{\partial 
g}{\partial p}
	\label{c:d}
\end{equation}
Here \( \kappa \) denotes the particle diffusion coefficient that is
assumed to be \( \kappa(p)=\kappa_0 p/p_0=\kappa_1 p/p_1 \), where \(
p_0 \) and \( p_1 \) are the injection and cut-off momenta,
respectively and the number density of the CRs is normalized to $4\pi g
dp/p $.  In the downstream region ($x <0$) we choose, as usual, a
spatially homogeneous solution given by \( U(x) \equiv -u_2 =const \)
and $g(x,p) \equiv g_0(p) $.  Thus \( u_2 \) is the (constant)
downstream speed.  Eq.(\ref{c:d}) governs the process of stationary
acceleration of the CRs that are drawn (injected) from the thermal
plasma at $p \sim p_0$ and leave the system at $p=p_1 \gg p_0 $.  Since
the pressure of accelerated particles directly influences the flow
profile $u(x)$ in the shock precursor, eq.(\ref{c:d}) should be
complemented by the conditions of the mass and momentum conservation
\begin{eqnarray}
	\rho u & = & \rho_1 u_1,
	\label{c:e}  \\
	P_{\rm c}+\rho u^2 & = & \rho_1 u_1^2 
	\label{ber}
\end{eqnarray}
Here \( \rho(x) \) is the mass density, \( \rho_1 =\rho(\infty )\),  
\(P_{\rm c} \) is the CR pressure
\begin{equation}
	P_{\rm c}(x)= \frac{4\pi}{3} mc^2 \int_{p_0}^{p_1}\frac{p 
	dp}{\sqrt{p^2+1}} g(p,x)
	\label{P_c}
\end{equation}
and no seed particles are present, i.e.  \( P_{\rm c}(\infty) =0\).  The 
particle momentum \( p \) is normalized to \( mc \).  Eq.(\ref{ber}) is 
written in the region $x>0$ where we have neglected the contribution of the 
adiabatically compressed cold gas confining our consideration to 
sufficiently strong shocks with \( M^2 \equiv \rho_1 u_1^2/\gamma P_{\rm 
g1} \gg (u_1/u_0)^{\gamma} \), where \( \gamma \) is the specific heat 
ratio of the plasma (see paper I for a detailed discussion of this 
approximation).

The subshock strength can be determined from the ordinary Rankine-Hugoniot 
condition for the gas
\begin{equation}
	r_{\rm s} \equiv \frac{u_0}{u_2}=\frac{\gamma+1}{\gamma-1+ 
	2M^{-2}R^{\gamma +1}}
	\label{c:r}
\end{equation}
where \( R= u_1/u_0 \).
\subsection{Invariant form of the solution}
It is quite clear that the formal solution to eq.(\ref{c:d}) can be
regularly found as a series in parameter $(u_1-u_0)/u_0$ if the latter
is small.  Fortunately (see paper I and \cite{m97b}) also in the case
of a very strong modification ($u_1 \gg u_0$) the structure of the
solution does not change significantly, provided that it is written in
appropriate variables.  The key variable here is simply the flow
potential
 \begin{equation}
	\Psi=\int_{0}^{x}u dx
	\label{f:p}
\end{equation} 
and the approximate solution in both cases can be 
represented as
\begin{equation}
	g=g_0(p)\exp \left\{-\frac{1+\beta}{\kappa}\Psi\right\}
	\label{c:d:sol}
\end{equation}
If the solution is efficient ($u_1 \gg u_0$), $\beta $ is given by the 
following relation
\begin{equation}
	\beta \equiv -\frac{1}{3}\frac{\partial \ln g_0}{\partial \ln p}
	\label{bet}
\end{equation}
In the case of inefficient solution ($u_1 \approx u_0$) the $\beta $-term 
in eq.(\ref{c:d:sol}) should be omitted altogether.  In fact, there is no 
much difference between these two cases as far as the formal representation 
of the solution is concerned; as it was shown in paper I, for efficient 
solutions $\beta \approx 1/6 $ over a broad momentum range.  On the other 
hand the explicit dependence of these solutions on $p$ and $x$ can be 
vastly different in these two cases since both the flow potential and the 
downstream spectrum may differ dramatically.  It is important to emphasize 
that it is this latter circumstance that constitutes the nonlinearity of 
the acceleration process, not the difference between the representations of 
the efficient and inefficient solutions by the formula (\ref{c:d:sol}) 
which is reflected in the \( \beta \) term.  The calculations in paper I 
show that this \( \beta \) term results in a numerical factor \( \theta 
\approx 1.09 \) in the expression for the shock compression ratio instead 
of \( \theta =1 \), that one would get without \( \beta \) term in the 
solution (\ref{c:d:sol}), i.e.  taking it in the form of inefficient 
solution.  Thus eq.(\ref{c:d:sol}) may be regarded as an invariant form of 
the approximate kinetic solution.  It creates a perspective of universal 
description of fairly different acceleration regimes.  For this purpose we 
extend the technique developed in paper I for describing the efficient and 
intermediate solutions to the case of inefficient solution.  Acting within 
the above-mentioned `ten percent' accuracy (\( \theta =1.09 \approx 1 \)), 
we will handle the \( \theta \)-factor somewhat loosely, putting \( \theta 
=1 \) in final results when we apply them to the inefficient part of the 
solution without a more accurate matching.  The price that we pay for the 
universality of description of all the three branches is not high, if one 
realizes that the global bifurcation properties of the system under 
question are not really known even qualitatively.
\section{Unified description of the acceleration process}
The basis of a universal treatment of acceleration solutions
is established in paper~I and its critical step is the derivation 
of an integral equation for a normalized spectral function
 \begin{equation}
 	J(p)=\frac{\bar V(p)}{\bar V(p_0)}
 	\label{J:d}
 \end{equation}
where \( \bar V \) is the spectral function introduced by the following 
relation
 \begin{equation}
 	\bar V(p)=\int_{0-}^{\infty} e^{-s(p)\Psi } du(\Psi)
 	\label{V:L}
 \end{equation} 
with 
 \begin{equation}
  	s(p)=\frac{1}{\kappa(p)\bar V(p)}\left[u_2+\bar V(p)+\frac{1}{3} 
	\frac{\partial \bar V}{\partial \ln p}\right]
  	\label{s}
  \end{equation}
The explicit dependence of the variable \( s \) on \( \bar{V} \) is not 
very critical here since \( \bar V \) effectively cancels out in 
eq.(\ref{s}) and eq.(\ref{V:L}) should be regarded logically as an integral 
transform \( u(\Psi ) \mapsto \bar V(p) \) rather than an equation for \( 
\bar V(p) \) given \( u(\Psi ) \).  An independent integral equation for \( 
\bar V(p) \) (or \( J \)) was, in turn, derived from the Bernoulli's 
integral (\ref{ber}).  Using a number of approximations discussed in paper 
I, this equation can be written down as follows
\begin{equation}
	J(\tau)=\frac{\zeta}{\varepsilon}\int_{\varepsilon}^{\varepsilon^{-1}}
	\frac{d\tau'}{\tau'+\tau}\frac{1}{\tau'J(\tau')}
	\exp\left[\frac{3}{\theta ( r_{\rm s} -1)}
	\int_{\varepsilon}^{\tau'}\frac{d\tau''}{\tau''J(\tau'')}\right] +1.
	\label{J}
\end{equation}
We have used the notations
\begin{equation}	
 	 \zeta =\nu R \exp \left[-\frac{3}{\theta (r_{\rm s}-1)} 
\int_{\varepsilon}^{1/\varepsilon }\frac{d 
\tau''}{\tau'' J(\tau'')}\right]
	\label{zet}
\end{equation}
\begin{equation}
	\tau=\frac{\kappa_0 s}{\varepsilon} \left(1-\frac{1}{r_{\rm 
s}}\right);\quad 
\varepsilon^2=\left(1-\frac{1}{r_{\rm s}}\right) \frac{p_0}{p_1} \theta 
\ll 1; 
	\label{tau:eps}
\end{equation}
We use here \( \nu \) as the injection rate that differs from the injection 
rate \( \eta \) used in paper I and is specified as follows
\begin{equation}
 	\nu \equiv  p_0 \eta \simeq\frac{p_0 r_{\rm s}}{r_{\rm s}-1} \frac{m 
n_{\rm c} 
c^2}{\rho_1 u_1^2} 
 	\label{inj:def}
 \end{equation}
where
\begin{equation}
	n_{\rm c}=4\pi \int_{p_0}^{\infty} g_0(p) dp/p
	\label{n_c}
\end{equation}
In order to close the system formed by eqs.(\ref{c:r}) and (\ref{J}) we
need another equation to relate the three variables \( (\nu, R, r_{\rm
s}) \) among whose only one, say \( \nu \) we consider as given.  As an
intermediate step we relate the precursor compression \( R \) and
the spectral function \( J \) by inverting eq.(\ref{V:L}):
\begin{equation}
 	\frac{du}{d\Psi} =\frac{1}{2\pi i}\int_{-i\infty}^{i\infty}e^{s\Psi} 
\bar V(s) ds +\Delta u \delta (\Psi )
 	\label{du}
 \end{equation}
where \( \delta \) is a delta function corresponding to the jump 
of \( u \) at the subshock. 
 Integrating then eq.(\ref{du}) over \( \Psi \) between \( \Psi =0+ \) and 
\( 
\infty \) we get
\begin{equation}
	u_1-u_0= \frac{\Delta u}{2\pi i}
\int_{-1/\varepsilon}^{-\varepsilon} \frac{d\tau}{\tau} 
\left[J(\tau+i0)-J(\tau-i0)\right]
	\label{u12}
\end{equation}
We have put \( \bar V(p_0) \approx \Delta u \) (see paper~I for  
details concerning this approximation).  The integral around the cut \( 
(-1/\varepsilon, -\varepsilon ) \) may be evaluated with the help of 
eq.(\ref{J}) and the last equation rewrites
\begin{equation}
	\frac{R-1}{1-r_{\rm s}^{-1}}=\frac{\zeta}{\varepsilon} U
	\label{R}
\end{equation}
where 
\begin{equation}
	U=\int_{\varepsilon}^{1/\varepsilon}\frac{d\tau}{\tau^2 
J(\tau)}e^{\Omega \phi (\tau )}
	\label{U}
\end{equation}
with
\begin{equation}
	\Omega =\frac{3}{\theta (r_{\rm s}-1)}
\quad {\rm and } \quad \phi (\tau )=\int_{\varepsilon}^{\tau} 
\frac{d\tau'}{\tau' J(\tau')}
	\label{om:fi}
\end{equation}
In general, the system formed by eqs.(\ref{c:r}), (\ref{J}), and (\ref{R}) 
may have multiple solutions.  On the other hand for sufficiently small 
injection rates \( \nu \) there must always be a solution to eq.(\ref{J}) 
that corresponds to the test particle acceleration regime in which \( J \to 
1 \) as \( \nu \to 0 \).  This solution can be written down in terms of a 
Neuman series in \( \nu \) or \( \zeta \) as follows
\begin{equation}
	J= 1+\frac{\zeta}{\varepsilon ^2} \ln \frac{\varepsilon^{-1} +\tau}
	{\varepsilon +\tau} + {\mathcal{O}}(\zeta^2)
	\label{JN}
\end{equation}
where we have put \( \theta =1 \) (see the preceding section) and \( r_{\rm 
s} =4 \) for simplicity.  Solution (\ref{JN}) is essentially perturbative 
and cannot describe two remaining solutions of eq.(\ref{J}) that appear 
beyond some \( \nu = \nu_1 > 0 \).  As we shall see all the three solutions 
may be conveniently described by the single valued function \( \nu = \nu 
(R) \) in the \( (R,\nu ) \) plane (Figure 1).  We term the solution with 
\( R > R_1 \) efficient, and that with \( R_2 < R < R_1 \) -- intermediate.  
It merges with the inefficient solution at the point \( \nu =\nu_2, R=R_2 
\).  For a fixed \( \nu \in (\nu_1,\nu_2) \) all the three solutions 
have different values of \( R \) and hence different subshock compression 
ratio \( r_{\rm s} \).

For the further bifurcation analysis the following 
relation appears to be useful.  It may be derived from eq.(\ref{J}) by 
multiplying both sides by \( (1/\tau^2 J)\exp(\Omega \phi )\) and 
integrating then the result between \( \varepsilon \) and \( 
\varepsilon^{-1} \):
\begin{equation}
	\frac{\zeta}{2 \varepsilon} U^2 +(1- \Omega) U+\varepsilon 
	e^{\Omega \phi_0}-\frac{1}{\varepsilon} =0
	\label{U^2}
\end{equation}
where
$$\phi_0 \equiv \phi(\epsilon^{-1})$$
Summarizing this subsection we note that eqs.(\ref{c:r}), (\ref{J}), and 
(\ref{R}) form a closed system for describing all the three branches of the 
solution.
\subsection{Branches 2 and 3}
As we mentioned the branches 2 and 3 of the function \( R(\nu) \) in
Figure 1 were studied in paper~I with the emphasis on the branch 3, that
corresponds to the efficient solution.  In this approximation \( J \gg
1 \) and hence one may neglect the second term on the right hand side
(r.h.s.) of eq.(\ref{J}) as well as the term in the exponent since \(
\zeta \gg \varepsilon \).  After the formal solution of the resulting
equation for \( J \) is obtained, eq.(\ref{zet}) for \( \zeta \)
serves as a nonlinear `dispersion relation' for this system.  There is a
pair of solutions in the region \( \nu > \nu_1 \) and there are no
solutions if \( \nu < \nu_1 \) within the approximation $\zeta \gg
\varepsilon $.  Note that the inefficient solution (branch one, $\zeta
\ll \varepsilon $) still exists and remains the only possible for \(
\nu < \nu_1 \).  The above approximation becomes better and better with
increasing \( \nu > \nu_1 \) for the branch 3 ($\zeta $ grows with $\nu
$) and it becomes worse for the branch 2, since in the latter case \(
\zeta \) decreases and the neglected terms become more and more
important.  We obtain here the part 2-3 of the bifurcation curve \(
R(\nu \)) in Figure 1 direct from eq.(\ref{U^2}) rather than from the
explicit solution of eq.(\ref{J}) as it was done in paper I.  The former 
procedure is more convenient for matching with the rest of this curve.  
First, neglecting the second and the third terms eq.(\ref{U^2}) simplifies 
to
\begin{equation}
	U\simeq \sqrt{\frac{2}{\zeta}}
	\label{Usim}
\end{equation}
The last approximation, being strictly valid for the branch 3, is also
good on some part of the branch 2 provided that (see
eqs.(\ref{zet},\ref{U^2}))
\begin{equation}
	\frac{\sqrt{\zeta}}{\varepsilon} \ga \left|1-\Omega\right|
	\label{cond3}
\end{equation}
\begin{equation}
	\sqrt{\zeta} \ga  \frac{1}{\ln 1/\varepsilon}
	\label{cond4}
\end{equation}
Inequality (\ref{cond4}) may be obtained estimating the integral in 
eq.(\ref{zet})
as \( \sqrt{2/\zeta} \) (paper I). These conditions are, generally 
speaking, opposite to the validity condition of the low \( \zeta 
\) 
expansion (\ref{JN}) (branch 1) that is
\begin{equation}
	 \sqrt{\zeta} < \varepsilon
	\label{cond1}
\end{equation}  
This makes the analytic description of the intermediate branch 2
(Figure 1), where \( \sqrt{\zeta} \) may approach \( \varepsilon \), more
difficult as it was explained earlier.  Fortunately, the factor \(
\left|1-\Omega\right| \) in eq.(\ref{cond3}) is rather small
numerically, unless \( r_{\rm s} \approx 1 \), and there exists an 
intersection between the half-intervals (\ref{cond3}) and (\ref{cond1}).  
The smallness of \( \left|1-\Omega\right| \) is ensured by the facts that 
\( \theta-1 \) is numerically small and \( r_{\rm s} \) is close to 4 on 
the branch 2 since \( R \) is appreciably smaller there than on the branch 
3 for the same \( M \) and \( \nu \).  The inequality (\ref{cond4}) is 
formally much more restrictive regarding the matching with the branch 1.  
Nevertheless, as we shall see in the next subsection, a quite accurate 
matching is still possible because the branch 1 can in fact be extended to 
the region (\ref{cond4}).  Besides that the third term in eq.(\ref{U^2}), 
whose neglecting requires the constraint (\ref{cond4}), will not 
significantly exceed, in fact, the last term even for \( R \ll R_1 \) (see 
eq.(\ref{fi_0}) below).  Coming back to eq.(\ref{R}) for the flow 
compression \( R \) we obtain
\begin{equation}
	R-1 =\frac{3\sqrt{2\zeta}}{4\varepsilon} \left(1-\frac{R^{8/3}}
	{M^2}\right)
	\label{R-1}
\end{equation}
We have used eq.(\ref{c:r}) with \( \gamma =5/3 \) and eq.(\ref{Usim}).
Now we are able to obtain the 2-3 piece of the  curve \( \nu = \nu (R) 
\). On the branch 3 and partly on 
the branch 2 not far from the critical point \( (\nu_1,R_1) \) we may 
replace the integral in eq.(\ref{zet}) by \( \sqrt{2/\zeta} \). 
Eq.(\ref{zet}) can thus be transformed to the required form
\begin{equation}
	\nu(R,M) =\frac{\zeta}{R}\exp\left(\frac{1}{\theta}\sqrt{\frac
	{2}{\zeta}}\frac{1+3R^{8/3}M^{-2}}{1-R^{8/3}M^{-2}}\right)
	\label{nu}
\end{equation}
where the function \( \zeta(R,M) \) should be taken from
eq.(\ref{R-1}).  The function \( \nu \) is plotted against \( R \) for
\( M = \infty \) and \( \delta \equiv p_0/p_1 = 10^{-4}\) in Figure 2 with
the dashed line.  As we mentioned this solution fails to work for \( R
\) sufficiently smaller than  \( R_1 \)  (Figure 1) and should be replaced
by a different solution that must be valid down to \( R=1, \nu=0 \).
The overlapping region to match both asymptotic solutions is presumably
within the interval \( R_2 < R <R_1 \).  For this purpose we obtain in
the next subsection an approximate solution that describes the branch 1
reasonably well, in particular it yields a qualitatively correct
behavior for \( \nu \to 0 \), is also valid for the branch 2 and may
thus be matched with the solution (\ref{nu}).
\subsection{Branches 1 and 2}
The key step here is to rewrite eq.(\ref{J}) as follows
\begin{equation}
	J(\phi )=\frac{\zeta}{\varepsilon}\int_{0}^{\phi_0}\frac{d\phi'}
	{\tau(\phi') + \tau(\phi)}e^{\Omega \phi'} +1
	\label{J2}
\end{equation}
where the function \( \tau(\phi) \) is defined by eq.(\ref{om:fi}).
Our nearest goal is to work out a usable approximation to the solution
of this equation that is good for the region of sufficiently small \( R
< R_1 \).  As we have seen in the
previous subsection, the branch 3 is characterized by the condition \(
\Omega \phi_0 \sim 1 \) for \( R \sim R_1 \) and  \( \Omega \phi_0 \ll
1\) for \( R \gg R_1 \), which facilitates the study of efficient
solution far from the critical point \( R=R_1 \).  Therefore it is
natural to use the approximation \( \Omega \phi_0 \gg 1 \) for \( R <
R_1 \) and to examine the possibility of matching the obtained
asymptotic result with the solution (\ref{nu}).  In this approximation
(\( \Omega \phi_0 \gg 1 \)) the integral in eq.(\ref{J2}) is dominated
by its upper limit and we may write it as
\begin{equation}
	J(\tau) = \frac{\zeta}{\Omega} \frac{e^{\Omega \phi_0}}
	{1+\varepsilon \tau}+1
	\label{J3}
\end{equation}
We will continue the discussion of the accuracy of this approximation for 
moderate values of \( \Omega \phi_0 \) in the next section arguing in terms 
of a very good matching of this solution at \( \Omega \phi_0 \sim 1 \) with 
the solution given in the preceding subsection for the 2-3 piece of the 
curve \( \nu (R) \).  Substituting now the asymptotic result (\ref{J3}) in 
eq.(\ref{om:fi}) we get
\begin{equation}
	\phi_0=\frac{\xi /\Omega}{\xi /\Omega+1}\ln \frac{\xi /\Omega+2}
	{\xi /\Omega+1}+\frac{2}{\xi /\Omega+1}\ln \frac{1}{\varepsilon}
	\label{fi_0}
\end{equation}
where 
\begin{equation}
	\xi= \nu R
	\label{xi:d}
\end{equation}
Next, since the flow compression \( R \) is not very large for 1-2
branches we can simplify the algebra by using the approximation of
infinite Mach number for these branches, i.e we put \( r_{\rm s}=4 \)
in eq.(\ref{U^2}).  This is a good approximation for \( R \ll
M^{3/4}\).  Neglecting also \(\theta -1 \approx 0.1 \) compared to \(
R-1 \ga 1 \), in eq.(\ref{U^2}), eqs.(\ref{zet}), (\ref{R}) and
(\ref{U^2}) yield
\begin{equation}
	(R-1)^2=\frac{9}{8}\xi \left(\frac{1}{\varepsilon^2}
e^{-\Omega \phi_0}-1 \right)
	\label{R1}
\end{equation}
The last approximation is formally inaccurate for \( R-1 \la \theta -1
\) as the inspection of eq.(\ref{U^2}) may show. On the other hand  
the qualitative behavior of \( \nu(R) \) remains unchanged in this region
and is essentially \( \nu \sim R-1 \). As we mentioned earlier
the value \( \theta \approx 1.09 \) was calculated in paper I for
the efficient (intermediate) solution  whereas for the inefficient
solution in the region \( R-1 \ll 1  \) the value \( \theta=1 \) is, in
fact, a better choice. In this case eq.(\ref{R1}) is virtually
also valid for small \( R-1 \) provided that \( r_{\rm s} \approx 4 \),
which is definitely the case for such a small $R $.  Substituting \(
\phi_0 \) from eq.(\ref{fi_0}) and taking the relation \( \Omega
\approx 1/\theta \) into account we finally obtain
\begin{equation}
	(R-1)^2=\frac{9}{8}\xi \left[\left(\frac{1+\theta \xi}{2+\theta 
\xi}\right)^{\frac{\xi}{1+\theta\xi}}\varepsilon^{2\frac{1/\theta -1 
-\theta\xi}{1+\theta\xi}}-1 \right]	
	\label{R2}
\end{equation}
Together with the relation (\ref{xi:d}) this equation determines the
1-2 part of the curve \( \nu=\nu(R) \). It is plotted in Figure 2 with the
solid line.  The both solutions may be matched smoothly by the
construction of an appropriate intermediate expansion. It is in fact
not needed for our purposes since the deviation of these two solutions
is numerically insignificant in an extended region between the
intersection points.
\section{The overall picture of stationary acceleration}
In the preceding section we have demonstrated that the method of
integral equation developed in paper~I allows one to describe the
process of steady shock acceleration on a universal basis, which gives
to the notions of efficient and inefficient acceleration a concrete
mathematical content in the form of a relevant bifurcation analysis.
Although different asymptotic approaches for describing
efficient and inefficient regimes are applied in the present study, the
both solutions match in the region of the intermediate acceleration
regime very well.  This is in fact an overlapping region for the two
solutions in the sense of the obtained asymptotic results.  Generally,
the integral equation (\ref{J}) describes all the three regimes on the
same basis so that the matching procedure can be improved and a
uniformly valid asymptotic expansion can be constructed. However the
achieved accuracy is sufficient for our purposes.

In fact we are already in a position to describe the   stationary
solution space in terms of system parameters and dependent variables
that characterize the acceleration process. In principle, our parameter 
space is
two-dimensional and contains the Mach number \( M \) and the cut-off
momentum \( p_1 \) (we will use the parameter \( p_1/p_0 \equiv
\delta^{-1} \gg 1 \)).  It should be reminded that our consideration is
restricted to the region \( M \gg R^{5/6}, \, p_1 \gg 1 \)
(Sec.\ref{ks}).  A convenient dependent variable is the flow
compression \( R \) that obviously signifies the efficiency of
acceleration. In the present study, however, we add to this parameter space 
also the
injection rate \( \nu \), since this latter, even though being in
principle calculable, may vary depending on the model of the subshock
dissipation used (see the corresponding discussion in Appendix A of
paper I).  Furthermore, as we have seen, the solution may be
conveniently represented in the form of a single valued function \( \nu
(R) \) that is shown in Figure 2.  Thus we perform our bifurcation
analysis here in three dimensional parameter space.  In a future work
we will reduce the parameter space to its natural dimensionality (two)
by specifying the injection rate given the subshock conditions.

In fact the character of bifurcations may be seen from the surface plots \( 
\nu = \nu(R,M) \) at fixed \( p_1 \) and \( \nu = \nu (R,p_1) \) at fixed 
\( M \), respectively.  These surfaces are shown in Figure 3a,b.  One sees 
that the multiplicity of the solution \( R = R(\nu ) \) is always present 
for sufficiently large values of \( M \) and \( p_1 \).  We have not 
plotted, however, the region of lower Mach numbers in Figure 3a where the 
solution indeed becomes unique for all \( \nu \).  The approximation \( M 
\to \infty \) adopted for describing 1-2 branches is inaccurate there and 
the usage of eq.(\ref{U^2}) is required for plotting the surface \( 
\nu(R,M) \) instead of a simplified version of it given by 
eq.(\ref{Usim}).  This is clearly indicated by a less accurate matching of 
the 1-2 and 2-3 asymptotic solutions for smaller \( M \) in Figure 3a.  The 
tendency to the uniqueness for lower \( M \) is quite obvious already from 
this figure.  A more accurate investigation of this region can be 
conveniently done by a numerical solution of eq.(\ref{J}), or better, of a 
more general equation in paper I which we leave for a separate study.

As it is seen from Figure 3a, the effect of the finite Mach number \( M \)
results in the fast increase of \( \nu (R,M) \) in the region \( R \la
M^{3/4} \), see also Figure 4.  Therefore, for sufficiently large $\nu = 
const $ where $M^{3/4} < \nu /\delta $, $R $ scales as $R \sim M^{3/4} $.  
In the opposite case, $M^{3/4} > \nu /\delta $ the compression ratio, 
represented as a function of $M $ saturates at the level $R \sim \nu 
/\delta $.  In this regime the behavior of the solution is close to that 
for \( M = \infty \).
\subsection{Critical injections.  The \tfm paradox 
revisited}\label{crit:inj}
The next issue we address here is the calculation of critical injections \( 
\nu_1 \) and \(\nu_2 \).  From eqs.(\ref{R-1}) and (\ref{nu}), restricting 
ourselves for simplicity to the case \( M \to \infty \) we 
calculate\footnote{The critical injection \( \nu_1 \) obtained here without 
an explicit usage of the solution to eq.(\ref{J}) differs from the 
correspondent result (eq.(70)) in paper I by a numerical factor \( \sim 1 
\).  This difference was explained in detail in paper I (see eq.(57) and 
the text below it).}
\begin{equation}
	\nu_1=2 e\sqrt{\frac{\delta}{3\theta}}; \quad R_1=\theta^{-3/2}
	\sqrt{\frac{3}{\delta}}
	\label{nu_1}
\end{equation}
To calculate \( \nu_2 \) we simplify eq.(\ref{R2}) as follows.  First, as 
discussed in Sec.2.2 we may set \( \theta =1 \).  We also assume \( \xi \ll 
1 \) but \( \delta^{-\xi} \gg 1\).  Then, from eq.(\ref{R1}) we have \( 
(R-1)^2 = \frac{9}{8}\xi \delta^{-\xi} \).  Substituting \( \xi =\nu R \) 
for the second critical injection \(\nu_2 \) and for \( R_2 \) we find the 
following results that we write with a `logarithmic accuracy'
\begin{equation}
	\nu_2 = \frac{2\sqrt{2}}{3}
\frac{\sqrt{\ln \ln \frac{1}{\delta}}}{\ln \frac{1}{\delta}};
\quad R_2=\frac{3}{2\sqrt{2}}\sqrt{\ln \ln \frac{1}{\delta}}
	\label{nu_2}
\end{equation}
Clearly, the nonlinearity of the acceleration process will be important 
when the CR pressure in front of the shock is comparable with the ram 
pressure.  Therefore, an estimate for \( \nu_2 \) can be obtained directly 
from the condition \( P_{\rm c} \simeq \rho_1 u_1^2 \).  On putting \( g_0 
\simeq p_0/p , \, (r_{\rm s}=4)\) into \( P_{\rm c} \) and using the 
relativistic form of the partial pressure for all $p$ in eq.(\ref{P_c}), we 
get \( \nu_2 \simeq 1/\ln(1/\delta) \). A similar 
approach to estimating the backreaction of injected and subsequently 
accelerated particles onto the gas flow has been also applied by 
previous authors (see \eg  \cite{dmv}).
The both critical injections (eqs.(\ref{nu_1}, \ref{nu_2})) are 
plotted versus \( \delta \) in Figure 5.  One sees from Figs.1 and 5 that 
the limit \( \nu \to 0 \) automatically means \( R \to 1 \) (ordinary gas 
shock) for all \( \delta > 0 \).  Since \( \nu_2 \) vanishes with \( \delta 
\) much slower than \( \nu_1 \) there is always a significant gap between 
these two critical injections, i.e.  an extended region of multiple 
solutions even for very small values of~\( \delta \).

It is interesting to trace the deformation of the bifurcation curve \( \nu 
(R) \) under the transition to the two-fluid limit.  Strictly speaking 
there is no kinetic prototype for the smooth two-fluid efficient solution 
(\( r_{\rm s} \equiv 1 \)) as it was argued earlier.  At the same time this 
is not linked directly with the fact that \( R \) fails to tend to unity as 
\( \nu \to 0 \) in the \tfm and we explain this last anomaly of this model 
first and shall return to the issue of the subshock smoothing later.  To 
simplify the algebra we restrict our consideration to the case \( M = 
\infty \) that is quite representative in this regard.  For instance, one 
gets \( R=7 \) from the \tfm in this case, also for \( \nu =0 \).  Let us 
perform the transition \( \delta \to 0+ \) (fluid limit) in the kinetic 
description.  According to eqs.(\ref{nu_1}, \ref{nu_2}) we have \( \nu_2 
\to \nu_1 \to 0 \) as \( \delta \to 0 \).  That means that \( \nu(R) \to 0 
\) uniformly in the segment \( [1,R_1] \).  The fact that \( R_1 \) itself 
tends to infinity results simply from our assumption \( M =\infty \) and is 
not important here (so does \( R_2 \) although extremely slowly).  We know 
from eqs.(\ref{R-1}, \ref{nu}) that \( R,R_1 < M^{3/4} \) for all \( \nu >0 
\) and \( R \to M^{3/4} \) as \( \nu \to \infty \), Figure 3a.  Coming back 
to the \( \delta \to 0 \) limit we see that the function \( \nu(R, \delta ) 
\to \nu_0(R) \).  Furthermore, \( \nu_0(R) \equiv 0 \) for \(R \in 
[1,R_1] \) and rises sharply at some \( R_0 \in [R_1, M^{3/4}) \).  
That means that if we now let the actual injection rate tend to zero from 
above, \( R \) will tend to \( R_0 \ge R_1\) instead of \( R=1 \) as is 
always the case in kinetic description for all \( \delta > 0 \).  As we 
have seen \( R_0 \) may be rather large whereas in the \tfm \( R_0 =7 \) 
which is only because of the complete subshock smoothing.  Thus, there is 
nothing surprising that \( R \to R_0 \neq 1 \) as \( \nu \to 0 \) in the 
two-fluid description.  From the viewpoint of the superior kinetic solution 
this transition reads \( \delta \to 0,\, \nu \to 0 \) and yields \( R \to 
R_0(M) \gg 1 \) for \( M \gg 1 \).  The fact that \( R_0 \) turns out to be 
equal 7 in the \tfm is not important in the present context, as we 
mentioned.  On inverting the sequence, i.e.  letting \( \nu \to 0,\, \delta 
\to 0 \) we clearly get \( R \to 1 \), but this is definitely beyond the 
\tfm since the transition \( \delta \to 0 \) has already been made during 
the derivation of the two-fluid equations.  The sensitivity to the order of 
the limit transitions is a signature of the singular dependence of the 
kinetic solution upon \( \delta \) at \( \delta =0 \) and has, at least 
formally, nothing to do with the TFM.

For \( M < \infty \) the above consideration needs some modification with 
an involved algebra but the main conclusion remains: since in the limit 
process \( \delta \to 0, \, \nu_2 > \nu_1 \) for sufficiently large \( M 
\), a correct behavior of \( R(\nu) \) at vanishing \( \nu \), i.e. 
the transition \( R \to 1 \) as \( \nu \to 0 \) is impossible within the 
TFM.

Let us look now into the issue of the subshock smoothing. It was shown in 
paper I that the subshock may indeed be reduced significantly and its 
strength for sufficiently large \( M  \) depends on only one parameter. 
This is a most important system parameter governing the nonlinear shock 
acceleration. It equals
\begin{equation}
	\Lambda \equiv \frac{\Lambda_1}{M^{3/4}} \equiv  
	\frac{\nu}{\delta \cdot M^{3/4}} 
	\label{Lam}
\end{equation}
The subshock strength, when it is weak, scales as (see eq.(83) in paper I)
\begin{equation}
	r_{\rm s} -1 \sim 1/\Lambda, \quad \Lambda \gg 1
	\label{r_s}
\end{equation}
Since the \tfm implies that \( \delta =0 \), there is again nothing 
strange that it produces a completely smooth solution \( r_{\rm s} =1 \).

Therefore the \tfm solution is completely understandable and in certain 
respects may serve as a limit of the full kinetic solution.  The problem is 
that it is a singular limit and it might thus formally be applied -- 
provided that the closure parameters are found -- only in a small (in fact 
measure zero) part of parameter space, namely where \( \Lambda^{-1} \equiv 
\delta \cdot M^{3/4}/\nu =0 \).  The kinetic solution differs dramatically 
from the \tfm solution even for \( 1 \ll \Lambda < \infty \).  This is 
perhaps the price we have to pay for describing an intrinsically 
kinetic problem quasi hydrodynamically.

Turning now to the situation of finite \( \Lambda \), i.e., \( p_1 < \infty 
\) which, as we just have seen, should be addressed kinetically, we note 
that the parameter \( p_1 \) may vary not only from one steadily 
accelerating shock to another but it can grow as a maximum momentum in a 
time dependent acceleration process and can thus pass through the 
critical values of the corresponding bifurcation diagram.  If this time 
evolution is sufficiently slow the steady state solution with slowly 
varying parameters should be a good approximation provided that the temporal 
behavior of these parameters is determined.  We continue our bifurcation 
analysis from this perspective in the next section.
\section{Possible scenario of time dependent acceleration} 
It is convenient to argue in terms of a familiar projection method (see
\eg \cite{jos} for its hydrodynamic version). We start from a general
evolution equation in a symbolic form that governs the acceleration
process
\begin{equation}
	\frac{d w}{d t}=W(w,\mu)
	\label{ev:eq}
\end{equation}
where \( w \) is a suitable vector of state of the system that contains 
both the particle distribution \( g(x,p,t) \) and the flow structure \( 
u(x,t) \).  Eq.(\ref{ev:eq}) implies the diffusion-convection equation 
along with the continuity and Navier-Stokes equations in appropriate 
representations.  We denoted as \( \mu \) the set of relevant parameters 
like \( M \) and \( \nu \).  As one of the projections of \( w \) (or \( u) 
\) we may choose \( R \equiv u_1/u_0 \) and the correspondent component of 
eq.(\ref{ev:eq}) takes the form of an ordinary differential equation (ODE)
\begin{equation}
 	\frac{d R}{ d t}= W_1(R,r_{\rm s},p_1, \ldots , \mu)
 	\label{R:pr}
 \end{equation}
Here we regard \( p_1 \) as a dependent variable like \( R \) and \( r_{\rm 
s} \) and others denoted by the ellipsis rather than a parameter that 
belongs to the set \( \mu \) which would be appropriate for a problem with 
the fixed cut-off momentum considered in the previous sections.  One may 
also write the similar equations for \( r_{\rm s } \), \( p_1 \) etc.
\begin{eqnarray}
	\frac{d r_{\rm s}}{d t} & = & W_2(R,r_{\rm s},p_1, \ldots , \mu)
	\label{r:pr}  \\
	\frac{d p_1}{dt} & = & W_3(R,r_{\rm s},p_1, \ldots , \mu)
	\label{p:pr}
\end{eqnarray}
What we have described in the preceding sections corresponds to \( W_3 
\equiv 0 \) as an external constraint and we have found in fact our steady 
state solutions as solutions of the system
\begin{equation}
	W_1=0, \quad W_2=0
	\label{}
\end{equation}
The last equation being resolved for \( r_{\rm s} \) simply yields the R-H 
relation (\ref{c:r}), whereas the first equation was studied in Sec.3 and 
for a fixed \( \nu \) it provides the solution \( R=R(p_1,M) \), 
represented \eg in Figs.3-4.  Now it is convenient to depict this solution 
for different values of \( \nu \) and for a fixed \( M \) as shown in 
Figure 6.  For \( p > p_{1\, \rm cr} \) and depending on \( \nu \) and \( 
p_1 \) either one or three solutions occur.  Let us fix \( \nu = c \) for 
which the three solutions coexist in the interval \( p_1^{(1)} < p_1 < 
p_1^{(2)} \), whereas for \( p_1 <p_1^{(1)} \) (\( p_1 > p_1^{(2)} \)) only 
the inefficient (efficient) solution occurs.  It is obvious that the time 
scale in eq.(\ref{r:pr}) is much shorter than that in eqs.(\ref{R:pr}) and 
(\ref{p:pr}) since \( R \) and \( p_1 \) variations are followed by the 
subshock strength \( r_{\rm s} \) almost instantaneously.  The latter may 
be then found from the equation \( W_2(R,r_{\rm s},p_1) =0 \) as a function 
of \( R \) and \( p_1 \) and be then substituted into eqs.(\ref{R:pr}) and 
(\ref{p:pr}).  The time scales in eqs.(\ref{R:pr}) and (\ref{p:pr}) are 
generally not well separated.  Nevertheless, since significant variations 
in \( R \) can take place under constant \( p_1 \) (not vice versa, 
however), we may assume tentatively that \( p_1 \) evolves slowly compared 
to the transition time between equilibria of eq.(\ref{R:pr}) that are given 
by \( W_1(R,r_{\rm s}(R),p_1)=0 \) and plotted schematically in Figure 6 
as \( R(p_1) \).

Relaxing the condition \( W_3 = 0 \) suggests the following scenario of the 
time dependent acceleration.  Suppose \( \nu = c =const \) and the cut-off 
momentum $p_1(t)$ increases in time beginning from a value that is in the 
interval $p_{1 {\rm \, cr}} < p_1 < p_1^{(1)}(\nu)$ (Figure 6).  As long as 
the lower branch $R_{\rm l} (p_1)$ is stable, the system will advance along 
this branch towards higher $p_1$.  When the second critical point 
$p_1^{(2)}(\nu)$ is reached, the flow profile together with the underlying 
particle distribution will restructure in such a way that a much more 
efficient acceleration regime corresponding to the upper branch $R_{\rm 
u}(p_1)$ emerges.  If $R_{\rm u}(p_1^{(2)}) < M^{3/4}$ the subshock will 
not be reduced significantly which is the necessary condition for keeping 
injection at the same level, and the acceleration process may continue 
towards even higher $p_1$ along the efficient branch $R_{\rm u}$.  It is 
interesting to note that if, after this transition $p_1$ starts to decrease 
instead, there will be a hysteretic behavior in the acceleration process as 
it is shown in Figure 6.  The decrease of $p_1 $ may be caused by an 
enhance of losses due to \eg parameter variations on the shock path.

If $R_{\rm u}(p_1^{(2)}) > M^{3/4}$ the transition to the upper branch will 
be accompanied by a strong subshock reduction and, as a result, by a 
considerable variation in $\nu $.  It should be noted, however, that if the 
subshock is still not very weak, say $r_{\rm s} \ga 2$, then $\nu $ not 
necessarily decreases with decreasing $r_{\rm s}$ (see \cite{mv95}).  This is 
mostly an effect of normalization of injection rate $\nu$, the number of 
injected particles does decrease since the spectrum becomes steeper at \( p 
\sim p_0 \) (see eq.(\ref{inj:def})).  Besides that, this consideration is 
based on a purely kinematic treatment of the leakage from the downstream 
medium and, in addition, is limited to relatively strong subshocks.  When 
$r_{\rm s}$ drops to $r_{\rm s} \ga 1$ the injection rate $\nu $ may also 
become very low.  This means that the solution must move to a lower level 
$\nu = const $ to remain quasi-steady.  Such a process will depend very 
much on the injection model in use and on the underlying subshock 
dissipation mechanism which are beyond the scope of the present paper (see 
\cite{m97c}, \cite{mv97}).  One may only hypothesize that if the subshock 
reduction is very strong and $\nu $ substantially decreases during this 
transition, the solution should either evolve being close to the critical 
injection rate \( \nu_1 \) or, if \( \nu \) drops essentially below \( 
\nu_1 \) the solution may return to the lower branch again to 
display thus essentially time dependent behavior.

The above consideration implies the stability of the branch \( R_l \) 
so that the stability of \( R_{\rm u} \) and instability of \( R_{\rm i} 
\) follow just from the continuity of \( W_1(R) \) in eq.(\ref{R:pr}). A 
different situation occurs when the branch \( R_{\rm i} \) is stable, at 
least for certain values of \( p_1 \), and therefore \( R_l \) and \( 
R_{\rm u} \) are both unstable for the same \( p_1 \).  Then, as \( p_1 \) 
grows, the solution evolves along \( R_{\rm i} \) up to the point \( 
p_1=p_1^{(2)} \), beyond which the acceleration process again cannot 
continue in a quasi steady manner.  In a similar situation occurring in the 
\tfm it is usually argued that the intermediate solution is unstable (DV).  
This is indeed typical for systems with S-type response curves like ours.  
We do not consider this question here in any detail but we note that an 
extended system of ODEs (eqs.(\ref{R:pr}, \ref{r:pr}, \ref{p:pr})) or 
even its further 
extensions will almost certainly possess a much more rich invariant 
manifold than the fixed points \( R_l, \, R_{\rm i} \) and \( R_{\rm u} \) 
of eq.(\ref{R:pr}).

Interestingly, on the branches \( R_l \) and \( R_{\rm u} \), \( R \) 
increases with \( M \) whereas it decreases with \( M \) on the branch \( 
R_{\rm i} \) (see \eg Figure 4).  Similarly, under fixed \( M \) and \( p_1 
\), \( R_i \) decreases with \( \nu \), again opposite to \( R_l \) and \( 
R_{\rm u} \) (Figures 2, 3b, 6).  This region of an anomalous behavior (like 
negative differential conductivity) may even be identifiable in steady 
state numerical simulations.  It should be noted, however, that for very 
large Mach numbers \( M > R^{4/3} \) the compression ratio becomes 
practically independent of \( M \) on the all three branches (see also 
paper I).  
\section{Discussion and conclusions}\label{disc}
The early works on the nonlinearly modified CR shocks inspired the hope 
that, because of a very high acceleration efficiency in the nonlinear 
regime, the overall CR production should not be very sensitive to 
injection and just a qualitative understanding of this complicated 
process suffices for quantitative calculations of 
acceleration efficiency.  The CR dominated (efficient) solutions of the 
two-fluid model (DV) strongly supported this idea.  In fact, such an 
optimism rests on the limited amount of energy available in
the gas flow to be converted into CRs.  Indeed, if we consider a strong shock ($ 
M \gg 1 $) we may calculate the acceleration efficiency, or the coefficient 
of the flow energy conversion, as $\varepsilon_{\rm conv} = P_{\rm 
c}(0)/\rho_1 u_1^2=1-1/R $, eq.(\ref{ber}).  Whenever the acceleration process 
is known to be in the nonlinear regime ($R \gg 1$), almost all the flow 
energy goes into CRs, practically independent of anything at all.

The main issue now is, under which circumstances the system may indeed be 
in a highly nonlinear acceleration regime.  As we have seen, the answer to 
{\em this} question depends {\em critically } on the injection rate: 
infinitesimal variations of $\nu $ in the vicinity of $\nu_1 $ or $\nu_2 $ 
can result in finite (and typically very large) variations of \( R \) due 
to transitions between different branches.  Also solutions belonging to the 
same branch are typically very sensitive to the injection rate \( \nu \).  
This may be easily understood from the inspection of \eg Figure 2 and 
Figure 3a where only on the inefficient branch \( R \) varies relatively 
slowly with \( \nu \), while in the cases of intermediate and especially 
efficient solution \( R \) changes very rapidly with injection.  There is 
of course an injection insensitive region belonging to efficient 
solution where the compression $R(\nu ) $ approaches its upper bound \( R 
\sim M^{3/4} \) at given $M$ (see Figure 3a, where this region may be 
identified as a very sharp growth of the function \( \nu(R,M) \) in the 
farthermost corner of the plot).  Such a behavior is caused by the 
requirement of a finite subshock.  It should be noted that this 
scaling has indeed been observed in some numerical works with fixed 
injection rates (\eg \cite{bky}).  However, this situation occurs at 
sufficiently large values of \(\nu \) and diminished \( r_{\rm s} \) and it 
is doubtful that such a high injection rate is possible at a weak subshock.  
For smaller \( \nu \), when \( \nu \ll \delta \cdot M^{3/4} \) (the system 
parameter \( \Lambda \ll 1 \)) an important signature of the stationary 
acceleration is that the compression ratio is practically independent of \( 
M \) (see paper I for further details).

Physically, the injection rate should be calculated selfconsistently using 
the solution of injection problem given subshock parameters.  The solution 
of this problem provides a function \( \nu = \nu_{\rm s }(R,M) \).  Then 
isolated solutions for \( R = R(M) \) might be obtained as intersection 
points of the curves \( \nu_{\rm s} \) and \( \nu \) (as shown in Figure 
1).  This solution may or may not be multiple depending on the character of 
the function \( \nu_{\rm s} \). In any case, the bifurcation diagrams 
alone do not suffice for determining the actual acceleration efficiency.  
The calculation of the injection $\nu_{\rm s} $ as a function of shock 
characteristics is equally important for this purpose. It should also be 
born in mind that the model considered here and in paper I gives 
an upper bounds to the actual acceleration efficiency. A number of not 
included factors may significantly decrease compression ratio \( R \), 
acceleration efficiency and the spectrum hardness (see paper I for a 
relevant discussion).

Turning to the time dependent acceleration we note that unless the 
scenario suggested in the previous section is totally unrealistic, a 
critical quantity that would determine the CR production is the cut-off 
momentum \( p_1^{(2)} \) beyond which the system jumps to the efficient 
acceleration regime. According to eq.({\ref{nu_2}) \(p_1^{(2)} \propto 
\exp (1/\nu)  \). Since in the Bohm limit we may write \( p_1(t) \propto 
t \), the corresponding critical time \( t_{\rm crit} \propto \exp 
(1/\nu)   \). In an accelerating object of a finite life time \( \tau \) 
(\eg supernova remnant, SNR) the main question, of course, is whether the 
condition \( \tau > t_{\rm crit} \) is fulfilled. This is again 
extremely sensitive to the injection rate.

Theoretically, the injection rate \( \nu_{\rm s} \) must not necessarily be 
as high as \( \nu_{2} \) for the acceleration process to become efficient.  
The condition \( \nu_{\rm s } > \nu_1 \) could suffice provided that the 
lower branch \( R_l \) (Figure 6) looses stability for \( p_1 \la p_1^{(1)} 
\).  Since for large values of \( p_1 \), \( \nu_2 \ga 10 \nu_1 \) (see 
Figure 5), this may determine the outcome of the acceleration process 
completely.  Another possibility to overcome the high \( \nu_2 \) threshold 
is an essentially stronger time dependence of the acceleration process 
than that discussed in the preceding section.  Basically, the bifurcation 
picture of this system is quite rich and promises an interesting dynamics.  
This is the more so as governing parameters are themselves subject for a 
temporal evolution.  They may change significantly during the acceleration 
process in a variety of astrophysical environments.  These variations may 
be of a quasi-external type like \eg decrease of the Mach number when the 
shock slows down.  Equally important may be an intrinsic variability 
associated with the growth of the maximum energy or with the heating 
of the upstream plasma by the CR driven turbulence (\cite{VDMcK}).  
Therefore, to comprehend the acceleration dynamics we must face the 
injection problem together with the physics of subshock dissipation and 
treat these problems selfconsistently with the above bifurcation analysis.
The fact that this system displays very much hysteresis emphasizes the 
necessity of this approach.

It is important to recognize that there is a serious drawback in the
way to a full calculation of the acceleration efficiency in concrete
astrophysical shocks, like \eg SNR shocks.  It originates from the
threshold nature of the acceleration process.  Indeed, the flow
structure changes quasi-abruptly when the critical injection $\nu_2 $
drops below $\nu_{\rm s}(R_2) $ as \( p_1 \) grows (or $\nu_{\rm
s}(R_1) $ becomes smaller than $\nu_1 $).  Since the function $\nu_{\rm
s} $ is very sensitive to local subshock conditions (a local
orientation of the magnetic field is perhaps the most obvious and very
important factor here), this transition occurs first at those parts of
the shock surface where $\nu_{\rm s} $ reaches its maximum.  This must
result in `hot spots' or `discharge' zones in the shock front where the
acceleration becomes efficient.  Then, the flow structure will be
essentially 3-dimensional (or at least quasi 2-dimensional), quite
complicated and probably unsteady.  The inhomogeneity of the ambient
medium (see \eg \cite{mcKee}) may very well result in a similar
effect.  Clearly, the one-dimensional calculations, even with a
properly determined injection rate \( \nu_{\rm s}(R,M) \) may give at
best an upper bound to the acceleration efficiency.  Even if the flow
remains quasi-laminar the overall efficiency will be reduced according
to the surface density of the hot spots on the shock front.  Besides
that the losses from the hot spots into the neighboring regions of
inefficient acceleration may significantly reduce the maximum energy. A
very important consequence of this would be the corresponding increase
of the critical injection, eq. (\ref{nu_1}) which may drive the system
below the threshold of the efficient acceleration.

\acknowledgments{I would like to thank Heinz
V\"olk for intersting  discussion. This work was done within the
Sonderforschungsbereich 328, ``Entwicklung von Galaxien'' of the
Deutsche Forschungsgemeinschaft (DFG).}

\newpage
	\figcaption[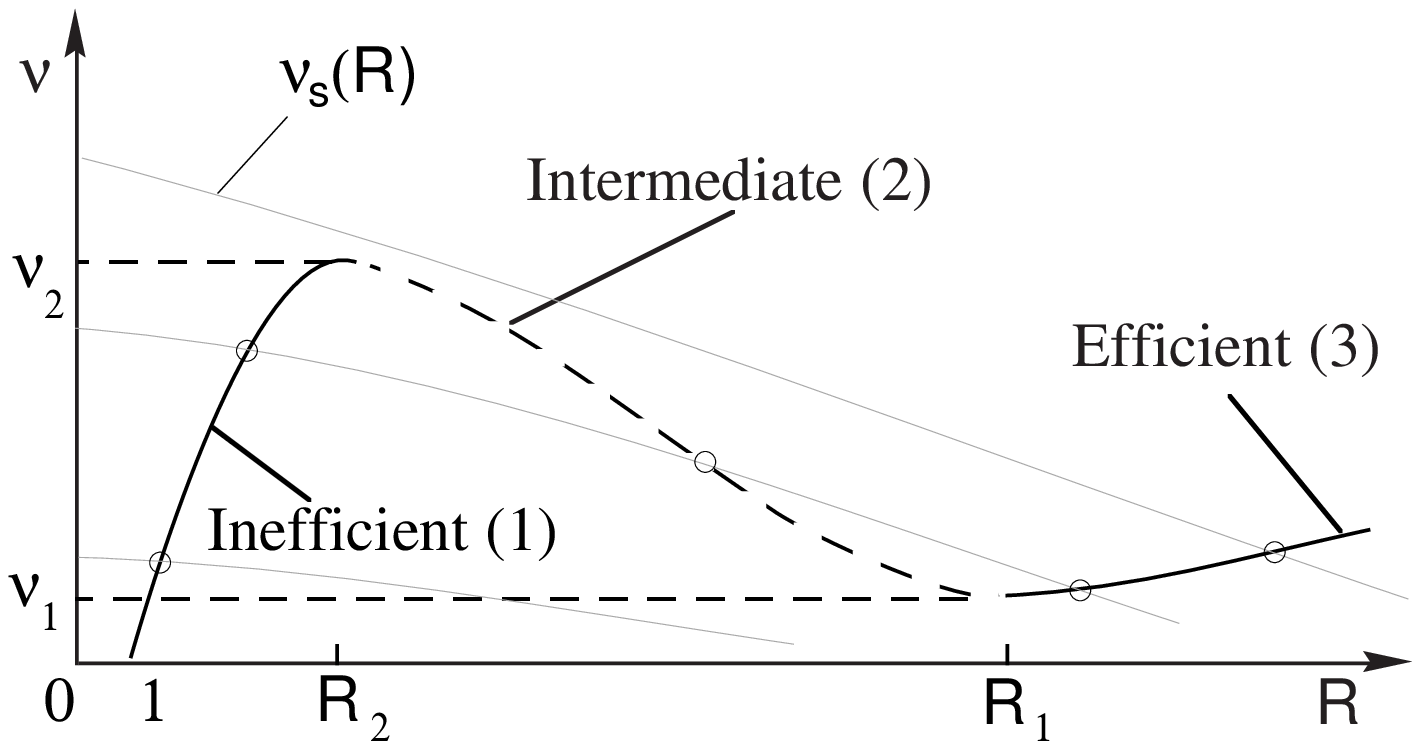]{The nonlinear response \( R \) 
of an accelerating shock to the thermal injection \( \nu \) represented
in the form of a single valued function \( \nu(R ) \).  Given \( \nu
\in (\nu_1,\nu_2 ) \) there are three substantially different
acceleration regimes.  A few possible graphs of \( \nu_{\rm s}(R) \)
(see Sec.\ref{disc}) are also drawn with the thin lines.  \label{fig1}}
	
	\figcaption[fig2.eps]{a.) Response curves calculated from
	eq.(\ref{nu}) (branches 2-3, dashed line) and from
	eq.(\ref{R2}) (branches 1-2, solid line), for \( M=\infty \)
	and \( \delta =10^{-4} \). 
	b.) Blow up of the peak in Figure 2a. \label{fig2}}

	\figcaption[fig3.eps]{a.) The surface of stationary solutions
\( \nu(R,M) \) plotted for \( \delta = 10^{-3} \).
b.) The same as in Figure 3a but \( \nu \) is given as a
function of \( R \) and \( \delta \) for \( M=\infty \). Both
surfaces are trimmed at high $\nu $. \label{fig3}}

	\figcaption[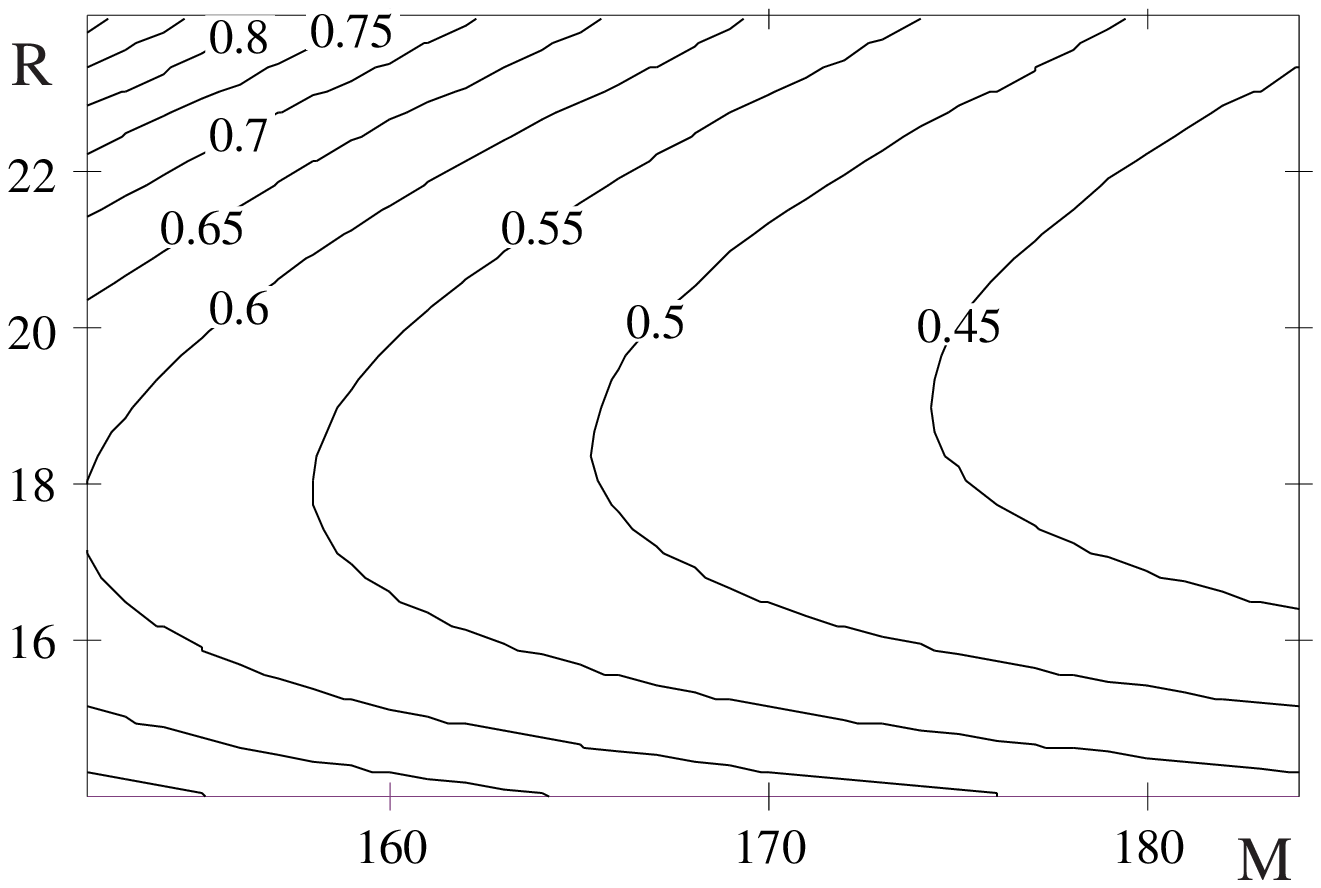]{The same as in Figure 3a but this time in
	the form of the contourplot \( \nu =const \) and for smaller \(
	M \), not shown in Figure 3a.\label{fig4}}

	\figcaption[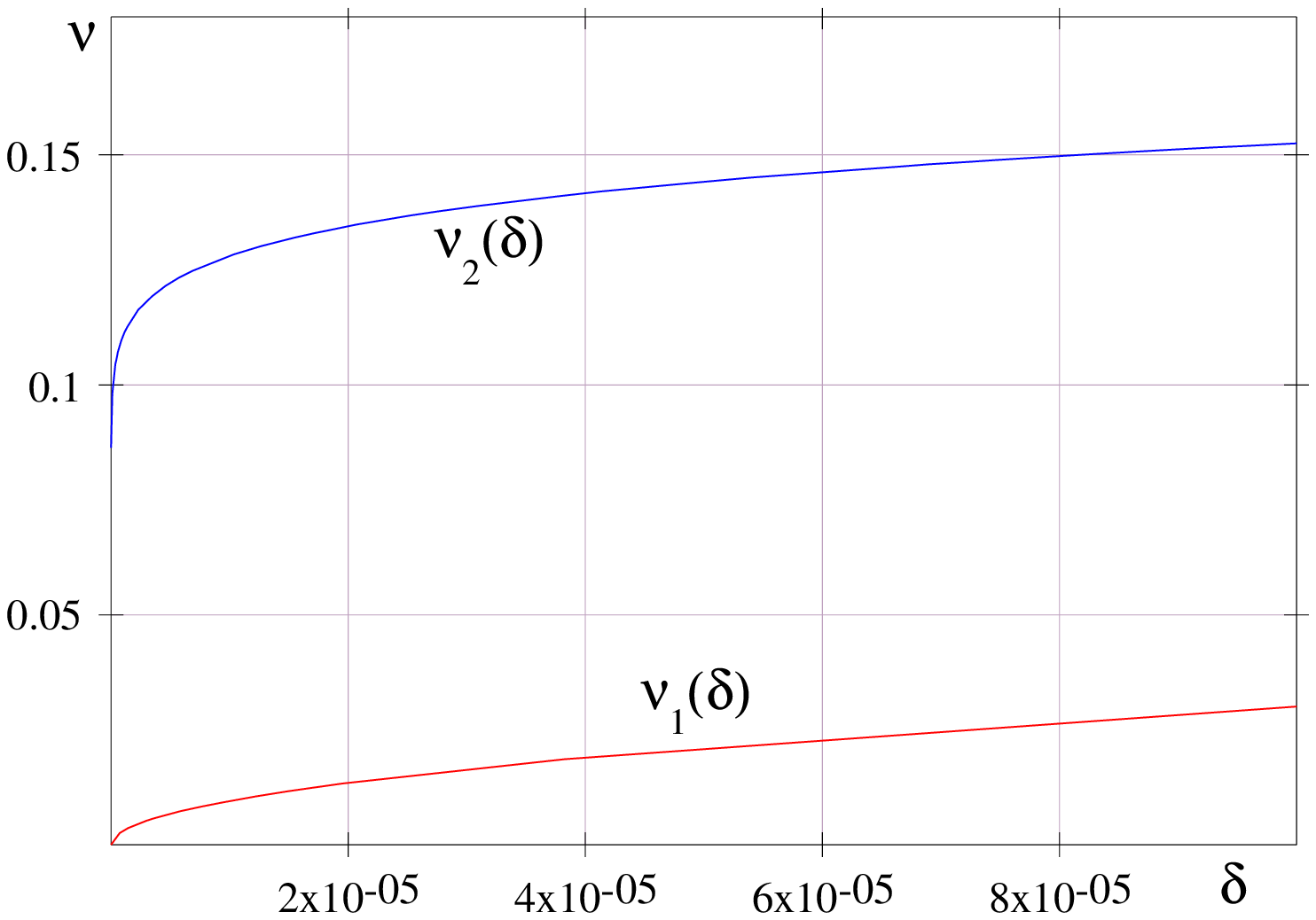]{Critical injections \( \nu_1 \) and \(
\nu_2 \) versus \( \delta \) given by eqs.(\ref{nu_1}) and
(\ref{nu_2}).\label{fig5}}

	\figcaption[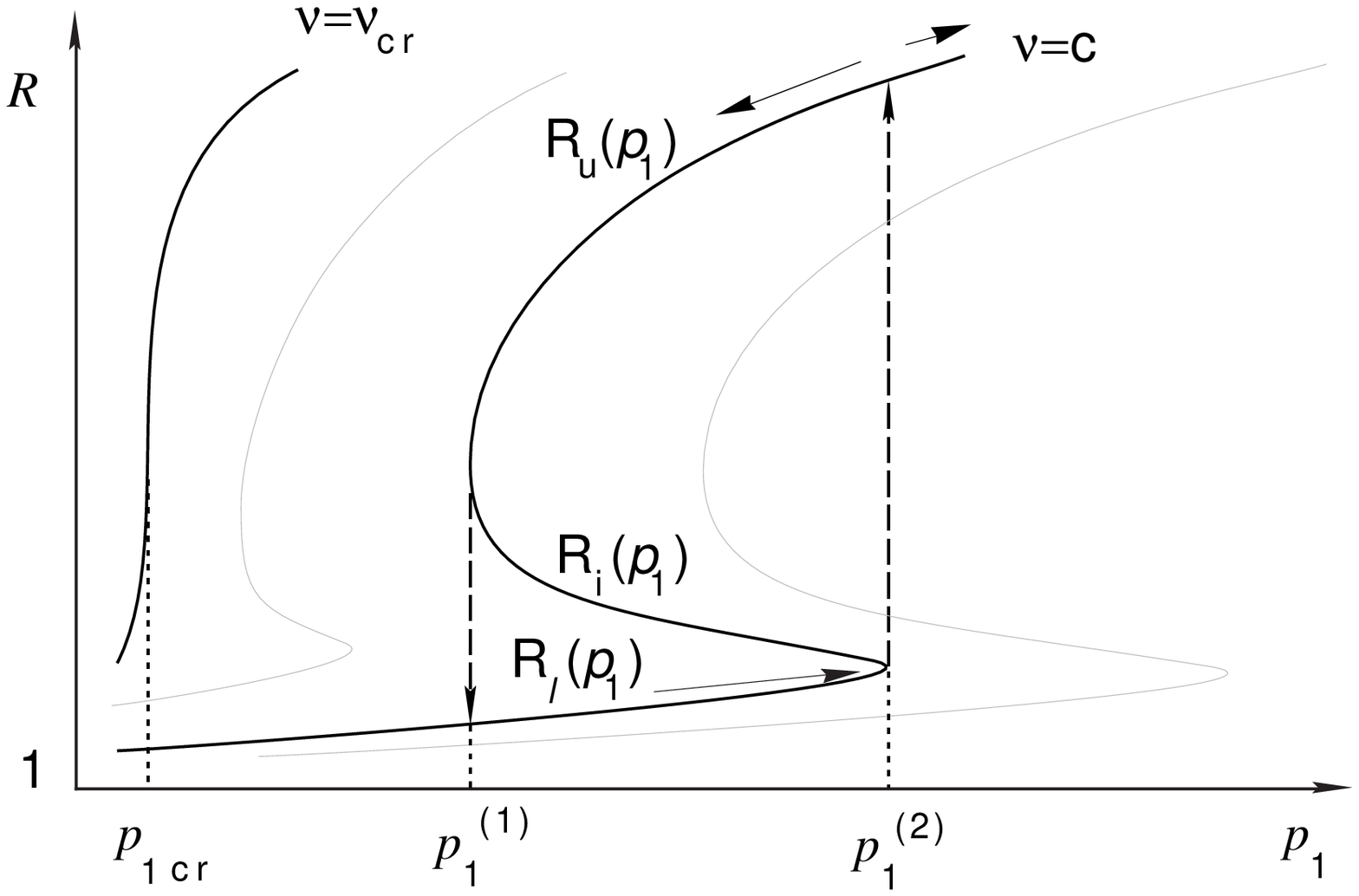]{Curves of constant \( \nu \) drawn
schematically on the basis of the surface plot shown in Figure
3b. The bifurcation of acceleration occurs when \( \nu \) crosses its
critical value, \( \nu_{\rm cr} \) (the solid curve marked by
$\nu=\nu_{\rm cr}$). For \( \nu < \nu_{\rm cr} \) and \( p_1 > p_{1\,
{\rm cr}} \) three different acceleration regimes emerge for \( p_1 \)
being within a certain interval (the solid curve marked by
$\nu=\nu_{\rm c}$).  These regimes, \( R_{l},\, R_i,\, R_{\rm u}(p_1)
\) correspond to the 1,2,3 branches in Figure 1. The two remaining
light curves correspond to $\nu $- values $c < \nu <\nu_{\rm cr} $ and
$\nu < c $.\label{fig6}}

\end{document}